\documentstyle[epsf,preprint,prb,aps]{revtex}
\def\d{\Delta}
\def\e{\epsilon}
\def\w{\omega}
\def\g{\Gamma}

\begin{document}
  \draft
\title{Incoherent tunnelling through two quantum dots with Coulomb interaction}
\author{P. Pals and A.MacKinnon}
\address{The Blackett Laboratory, Imperial College,
London SW7 2BZ, United Kingdom}
%\date{}
\maketitle

\begin{abstract}
The Ohmic conductance and current through two quantum dots in series 
is investigated for the case of incoherent tunnelling. A generalised 
master equation is employed to include the discrete nature of the 
energy levels. Regions of negative differential conductance can occur in
the I-V characteristics.
Transport is dominated by matching energy levels, even when they do not 
occur at the charge degeneracy points. 
\end{abstract}

%*************************************************************************

\narrowtext

\noindent
Due to the improvement of lithographical techniques on a nanometre scale 
in recent years it has become 
possible to study systems that were inaccessible to experimentation before.
The most easily controlled mesoscopic systems are defined in the 
2-dimensional electron gas of a semiconductor. By the application of 
gate electrodes \cite{Wees,Wharam} it is possible to confine the electron gas 
effectively to produce quantum wires and dots. 
This allows charge quantisation to be observed in the form of the Coulomb 
blockade \cite{Scott-Thomas,HoutBeen}. Moreover,  
the importance of size quantisation has been shown by Reed {\it et al.} 
who discerned discrete states in small quantum dots \cite{Reed}.  

The effect of the charge quantisation has been studied both for a single
dot \cite{Amman2} and a double dot \cite{Ruzin,Kemerink}. In both cases the 
level spectrum could be considered continuous. 
The effect of level quantisation on the Ohmic conductance through a 
single dot has been studied in the limit of weak coupling to the reservoirs.
Coherent \cite{Lee} and incoherent methods \cite{Been44} lead to the same
result in this limit. 

This paper will focus on the transport properties of a double dot 
with discrete equally spaced energy levels. The limit of 
completely incoherent tunnelling is studied, where the phase-breaking rate
is large and the tunnelling process is sequential. This justifies the use of 
a semi-classical method like the master equation. 
It must be noted that the phase-breaking time is still large compared to 
the time it takes for an electron to traverse the dot. This ensures that
the energy levels are quantised, due to the coherence of the wavefunctions 
inside the dot.

\section{Generalised master equation} \label{quant_master}
{
\noindent
When it is assumed that the tunnelling rates are small compared to the 
Coulomb energy and the average level spacing, then use of the master 
equation is justified \cite{Averin2}.
In order to take account of the discrete nature of the energy-levels in the 
dot, the master equation method for dots with a continuous energy level 
spectrum \cite{Amman2}
has to be generalised. The tunnelling rates not only depend on the 
number of electrons in the dot, but also on their configuration, i.e. how 
the electrons are distributed over the available energy levels. When there is 
a high relaxation rate, then the electrons will revert to their local 
equilibrium distribution between tunnelling events. This is likely to be the 
case for small tunnelling 
barriers which cause the electrons to spend a long time in the dot.  
Since this distribution will depend only on the temperature,
the state of the dot may yet again be described simply in terms of the 
number of electrons present in the dot.

Let $P(p,N)$ be the probability that the system is in a state which is 
characterised by $N$ electrons in the dot which are distributed over the 
energy levels according to a configuration $p$. 
Define $T_{L1}^{pp'}(N)$ as the tunnelling rate coefficient corresponding
to the transition $\{p,N+1\} \rightarrow \{p',N\}$ by an electron tunnelling 
through the left barrier. $T_{1L}^{p'p}(N)$ refers to the rate that corresponds
to the reverse process. The tunnelling rates through the right barrier 
are defined similarly. Finally, there are the relaxation rates $T_r^{pp'}(N)$ 
which account for the intra-dot transitions at a fixed electron occupation
$N$. 
The rate of change of the probability of occurrence of the state $\{p,N\}$ is
thus given by
\begin{eqnarray}
{d \over dt} P(p,N) &=& \sum_{\Omega=L,R} \sum_{p'} 
\left[T_{\Omega 1}^{p'p}(N-1) P(p',N-1)+
T_{1\Omega}^{p'p}(N) P(p',N+1) \right]    \nonumber \\
&-& \sum_{\Omega=L,R} \sum_{p'} \left[T_{\Omega 1}^{pp'}(N)-
T_{1\Omega}^{pp'}(N-1) \right] P(p,N)
\nonumber \\
&+& \sum_{p'} \left[T_r^{p'p}(N) P(p',N)-T_r^{pp'}(N) P(p,N) \right]
\end{eqnarray}
If the system consists of several dots in series, it is necessary to 
define the tunnelling rates between the dots. 
Define $T_{i,i\pm1}^{pp',qq'}(N_i,N_{i\pm1})$ as the tunnelling rate 
corresponding to the transition 
$\{p,N_i+1\} \rightarrow \{p',N_i\}$ and 
$\{q,N_{i\pm1}\}   \rightarrow \{q',N_{i\pm1}+1\}$
where the subscript $i$ labels the site of the dot. 
The evolution of the multi-particle state occupation probabilities 
for a dot not neighboured by a reservoir is given by
\begin{eqnarray}
{d \over dt} P(p,N_i) &=& \sum_{N_{i\pm1}} \sum_{p',q,q'}
 T_{i,i\pm1}^{p'p,qq'}(N_i,N_{i\pm1}) P_i(p',N_i+1) P_{i\pm1}(q,N_{i\pm1}) 
\nonumber \\
&+& \sum_{N_{i\pm1}} \sum_{p',q,q'} 
T_{i\pm1,i}^{qq',p'p}(N_{i\pm1},N_i-1) P_i(p',N_i-1) P_{i\pm1}(q,N_{i\pm1}+1) 
\nonumber \\
&-& \sum_{N_{i\pm1}} \sum_{p',q,q'} T_{i,i\pm1}^{pp',qq'}(N_i-1,N_{i\pm1})
		  P_i(p,N_i) P_{i\pm1}(q,N_{i\pm1}) \nonumber \\
&-& \sum_{N_{i\pm1}} \sum_{p',q,q'} T_{i\pm1,i}^{qq',pp'}(N_{i\pm1},N_i)
P_i(p,N_i)  P_{i\pm1}(q,N_{i\pm1}+1) \nonumber \\
&+& \sum_{p'} \left[T_r^{p'p}(N) P_i(p',N)-T_r^{pp'}(N) P_i(p,N) \right]
\end{eqnarray}
For both the single and the multiple dot system the steady state current can 
be written as
\begin{equation}
J = e \sum_{p,p'} \sum_{N} \left[T_{L1}^{pp'}(N) P(p,N) -T_{1L}^{p'p} P(p',N)
\right]
\end{equation}
\noindent
When it is assumed 
that the relaxation rates are large compared to the tunnelling rates, then 
the number of electrons in the dots is the only significant variable, since 
knowledge of this quantity
allows one to deduce the 
probabilities of the various electronic configurations. Hence the formalism 
is greatly simplified. However, the 
tunnelling rates between states with a different occupation number have to be
redefined to include the weighted sum over all possible tunnelling paths. 
Define $T_{L1}(N)$ and $T_{1L}(N)$ as the tunnelling rates for an electron
tunnelling into and out of the dot through the left barrier with $N$ other 
electrons present in the dot.
\begin{eqnarray}
T_{L1}(N) &=& \sum_{p,p'} T_{L1}^{pp'}(N) P(p|N) \\
T_{1L}(N) &=& \sum_{p.p'} T_{1L}^{pp'}(N) P(p|N+1)
\end{eqnarray}
where $P(p|N)$ is the conditional probability that the system is in 
configuration $p$ given that there are $N$ electrons in the dot. 
Since the master equation takes only single electron tunnelling events into 
consideration, the only contributions that need to be taken into account are 
those
where $p$ and $p'$ differ by the occupation of one energy level. Therefore 
the double sum over $p$ and $p'$ can be replaced by a single sum over the 
single particle energy levels $k$. 
\begin{eqnarray}
T_{L1}(N) &=& \sum_k T_{L1}^k(N) [1-P(k|N)] \\
T_{1L}(N) &=& \sum_k T_{1L}^k(N) P(k|N) 
\end{eqnarray}
Here $P(k|N)$ is the conditional probability that level $k$ is occupied 
given that there are $N$ electrons in the dot. 
For inter-dot tunnelling events the rates are re-expressed in terms of a 
sum over the tunnelling paths from
all levels $k$ in dot $i$ to all levels $l$ in dot $i\pm1$ 
\begin{equation}
T_{i,i\pm1}(N_i,N_{i\pm1}) = \sum_{k,l} T_{i,i\pm1}^{k,l}(N_i,N_{i\pm1})
P_i(k|N_i+1)[1-P_{i\pm1}(l|N_{i\pm1})]
\label{Tgen}
\end{equation} 

\noindent
For the case of negligible energy level broadening the probability 
$P_i(k|N_i)$ of finding an electron on level $k$ of dot $i$ given $N_i$ 
electrons can be determined from the Boltzmann distribution. 
\begin{eqnarray}
P_i(k|N_i) &=& e^{-\beta \e_k(N_i-1)} {Z_i^k(N_i-1)\over Z_i(N_i)} \label{p1} \\
1-P_i(k|N_i) &=&  {Z_i^k(N_i)\over Z_i(N_i)} \label{p2} 
\end{eqnarray}
where $\beta=1/k_BT$ and $\e_k(N_i-1)$ is the energy required to add an electron 
at an energy level $k$ when $N_i-1$ electrons are already present in the dot. 
$Z_i(N_i)$ is the partition function for $N_i$ electrons in dot $i$, 
$Z_i^k(N_i)$ is the conditional partition function for $N_i$ electrons 
given that level $k$ is unoccupied. 
When the conditional probabilities are substituted in equation \ref{Tgen}, 
one obtains
\begin{equation}
T_{i,i\pm1}(N_i,N_{i\pm1}) =  \sum_{k,l} {e^{-\beta \e_k(N_i)} Z_i^k(N_i) 
Z_{i\pm1}^l(N_{i\pm1}) \over Z_i(N_i) Z_{i\pm1}(N_{i\pm1})}T_{i,i\pm1}^{k,l}
(N_i,N_{i\pm1})
\label{Tinel}
\end{equation}
Tunnelling events between dots will in general not preserve energy, since it is
unlikely that energy levels will line up. 
The high relaxation rate suggests that all tunnelling events can be
considered inelastic. 
Hence
\begin{equation}
T_{i\pm1,i}^{l,k}(N_{i\pm1},N_i) = e^{-\beta[\e_l(N_{i\pm1})-\e_k(N_i)]}
T_{i,i\pm1}^{k,l}(N_i,N_{i\pm1})
\end{equation} 
It follows from the two preceding equations that 
the tunnelling rates are inelastic in the total free energy of the system.
This is also true for tunnelling between dot and reservoir. This implies that 
the following condition holds true at zero bias for $\cal N$ dots for all 
possible occupation numbers $N_i$.
\begin{equation}
T_{L1} \left( \prod_{i=1}^{{\cal N}-1} T_{i,i+1} \right) T_{{\cal N}R} =
T_{R{\cal N}} \left( \prod_{i=1}^{{\cal N}-1} T_{i+1,i} \right) T_{1L} 
\end{equation}
}

\section{The canonical distribution} \label{canon}
\begin{figure}
     \center{\hspace{0cm}
         \epsfxsize = 5in
         \epsffile{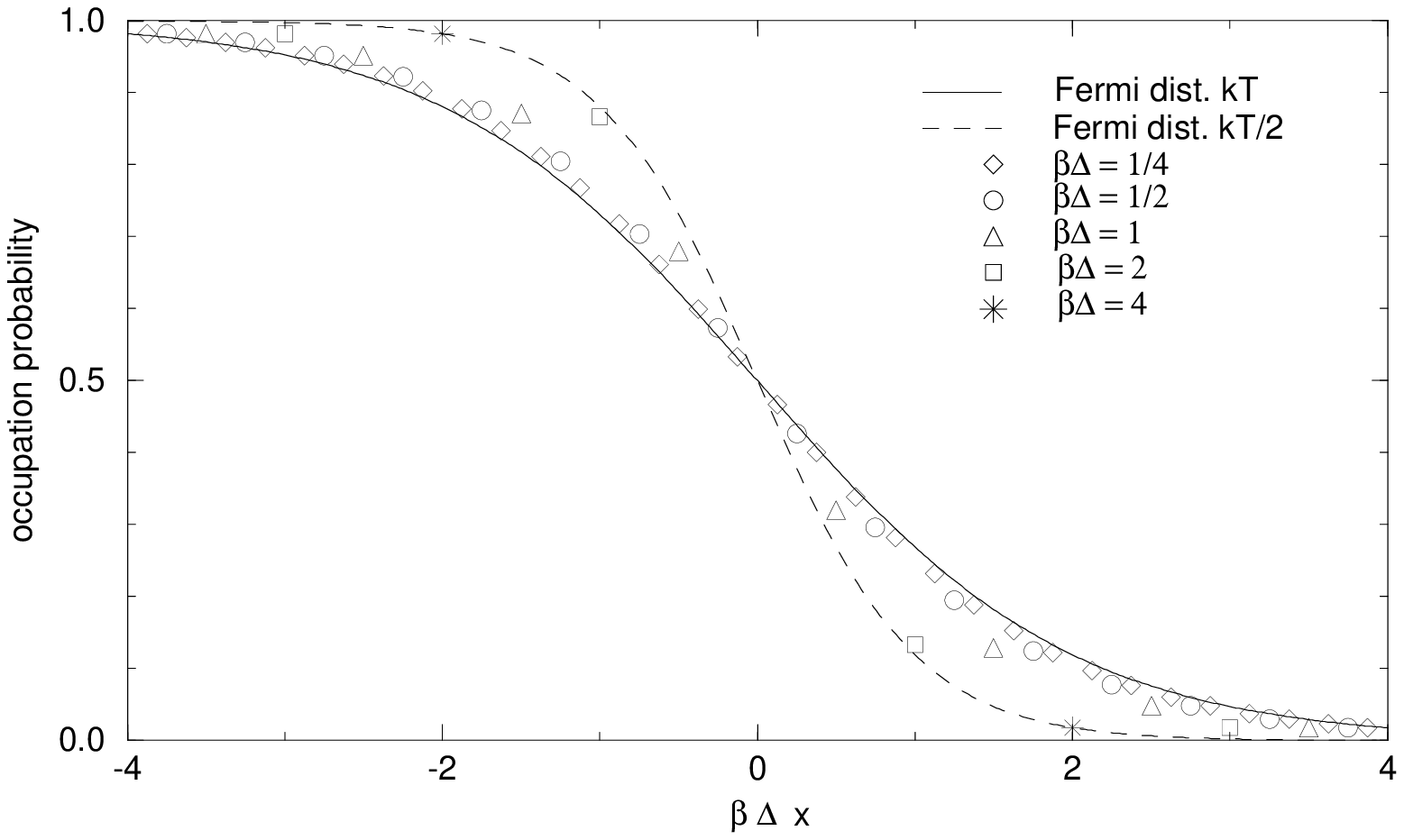}}
    \caption{Distribution function for a dot with equally spaced
      energy levels}
    \label{fig:dist}
\end{figure}

{
\noindent
One of the quantities that is needed to calculate the tunnelling rate 
between dots of specified occupation number is the probability $P(k|N)$ that
an energy level $k$ is occupied given a total occupation of $N$ electrons. 
When no relaxation of the electrons in the dot is allowed the occupation of 
the levels will be determined by the coupling to the reservoirs. However, 
in the presence of thermalisation the levels will be filled according to 
some equilibrium distribution independent of the energies at which the 
electrons initially entered the dot. 
In metals the energy-levels are filled according to the Fermi-Dirac 
distribution function. Naively one might expect this also to hold for 
discrete energy levels and this has been assumed in various papers. 
Unfortunately, this is not generally the case. In metallic 
systems the levels are very closely spaced so that the addition of an 
additional electron will have no effect on the distribution, whereas
this is not true for a discrete level spectrum.
In this section the correct
distribution will be calculated for a dot with an infinite number of 
equally spaced energy levels, containing $N$ electrons. The calculation is 
done for energy levels with negligible broadening ($\g \ll k_BT$).

Consider the partition function $Z(N)$ for a dot containing $N$ electrons,
with energy levels $\e_k = \e_0 + k \d_{\e}$. 
\begin{equation}
Z(N) = \sum_{p\{N\}} e^{-\beta(n_1 \e_1 + n_2 \e_2 + ...)}
\end{equation}
where the sum is taken over all realisations $p\{N\}$ with $N=\sum_k n_k$.
Explicitly expand out the contribution of the lowest energy level
(either $n_1 = 1$ or $n_1 = 0$).
\begin{equation}
Z(N) = e^{-\beta \e_1} \sum_{p\{N-1\}} e^{-\beta(n_2 \e_2 + n_3 \e_3 + ...)}
+ \sum_{p\{N\}} e^{-\beta(n_2 \e_2 + n_3 \e_3 + ...)}
\end{equation}
When it is assumed that there is an infinite number of energy levels, 
then the sum over the occupation numbers $n_k$ ($k > 1$) can be transformed 
to the partition function over all levels (including $k=1$). 
Rewriting $\e_k = \e_{k-1}+\d_{\e}$ and relabelling $n_k \rightarrow n_{k-1}$,
the following recursion relation is obtained. 
\begin{equation}
(1-e^{-\beta N \d_{\e}}) Z(N) = e^{-\beta \e_0} e^{-\beta N \d_{\e}} Z(N-1)
\end{equation}
With the boundary condition given by $Z(0)=1$, this is solved to give 
\begin{equation}
Z(N) = e^{-\beta N \e_0} \prod_{n=1}^N \left( {1\over e^{\beta n \d_{\e}}-1}
\right) 
\end{equation}

\noindent
This expression for the partition function does not 
explicitly include the many-body 
interaction. This 
is irrelevant in this context, since the number of electrons is kept
constant. However, when one wants to calculate the 
probability that the dot contains a certain number of electrons, the 
Coulomb interaction obviously needs to be included. For a 
simple model of the interaction where all pairs of electrons have an 
associated Coulomb repulsion $U$, the full interactive partition 
function is obtained
by multiplication with a factor $\exp[-\beta N(N-1)U/2]$. 

Now it is possible to do a similar calculation to obtain an expression for 
$Z^k(N)$, the conditional partition function for $N$ electrons 
given that level $k$ is unoccupied. 
 However, in contrast to the full partition function, this requires 
one to solve a recursion relation in both parameters $k$ and $N$. This can 
be avoided by using equations \ref{p1} and \ref{p2} to yield a 
recursion relation of the conditional probability $P(k|N)$ directly.
\begin{equation}
P(k|N) = e^{-\beta \e_0} e^{-\beta \d_{\e} k} {Z(N-1)\over Z(N)} 
\left[1-P(k|N-1)\right]
\end{equation}
Since the full partition functions are known, this is easily solved to give
\begin{equation}
P(k|N) = \sum_{q=1}^N (-1)^{q-1} \prod_{p=1}^q e^{-\beta \d_{\e} k} 
\left(e^{\beta \d_{\e} (N+1-p)}-1 \right)
\end{equation}

\noindent
In order to compare this with the Fermi-Dirac distribution function, one 
needs to consider a very large number of electrons in the dot. The 
occupation probability will depend only on the difference $x=k-N-1/2$.  
\begin{equation}
P_{\infty}(x) = \sum_{q=1}^{\infty} (-1)^{q-1} e^{-\beta \d_{\e} q (x+q/2)}
\end{equation}
For computational reasons the above equation is rewritten as
\begin{equation}
P_{\infty}(x) = 2 e^{{\beta \d_{\e} \over 2}(x^2-1/4)} 
 \left\{ \begin{array}{ll}
        \sum_{q=0}^{\infty} e^{-2\beta \d_{\e}(q-x/2+1/4)^2} 
          \sinh\left[\beta \d_{\e} (q-x/2+1/4) \right] & x<0 \\
        \sum_{q=0}^{\infty} e^{-2\beta \d_{\e}(q+x/2+3/4)^2} 
          \sinh\left[\beta \d_{\e} (q+x/2+3/4) \right] & x>0 
         \end{array} \right. 
\end{equation}

\noindent
The distribution function has been plotted in figure \ref{fig:dist}
for various values of $\beta \d_{\e}$. In the metallic limit, when the 
level spacing is small compared to the temperature, the distribution 
tends towards the Fermi-Dirac distribution function, as expected. 
When the temperature is comparable or smaller than the 
level spacing, the distribution deviates significantly. Its limiting
behaviour is well described by another Fermi-Dirac distribution function
with an effective temperature which is half the real temperature.  
A slightly more accurate result for the occupation probability in the 
limit $\beta \d_{\e} \gg 1$ is given by (note that $x \neq 0$ by 
definition)
\begin{equation}
P_{\infty}(x) = \left\{ \begin{array} {ll}
   {1\over e^{\beta \d_{\e}(x-1/2)}+1} & x<0 \\
   {1\over e^{\beta \d_{\e}(x+1/2)}+1} & x>0 
     \end{array} \right.
\end{equation}
}

\section{Tunnelling through a single dot} \label{quant_one}
\begin{figure}
     \center{\hspace{0cm}
         \epsfxsize = 5in
         \epsffile{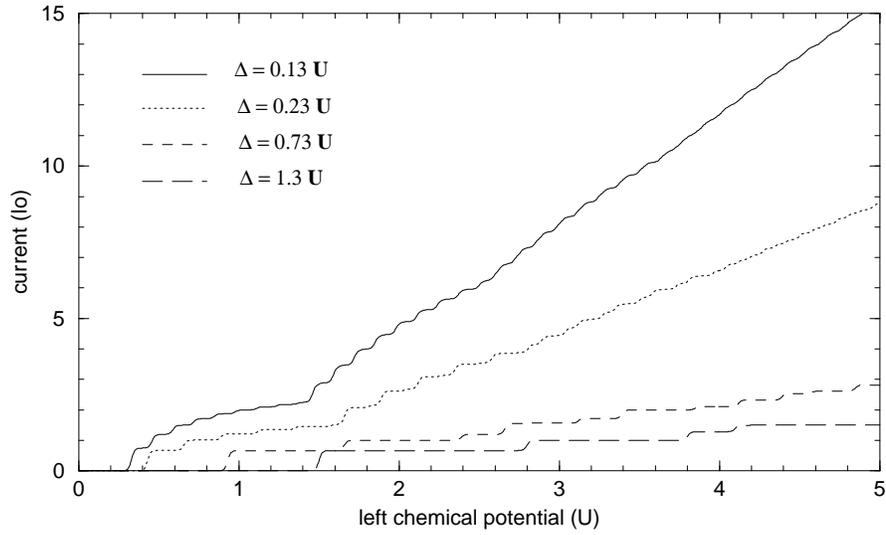}}
    \caption[I-V characteristics for various level spacings]
            {I-V characteristics for various level spacings ($\g_L=\g_R$)}
    \label{fig:disc1}
\end{figure}

\begin{figure}
     \center{\hspace{0cm}
         \epsfxsize = 5in
         \epsffile{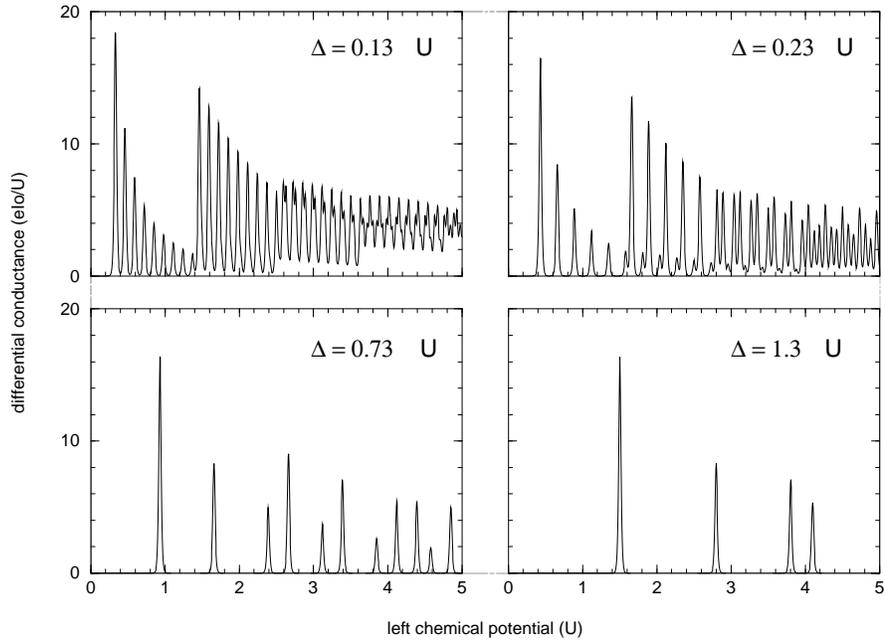}}
    \caption[Differential conductance for various level spacings]
            {Differential conductance for various level spacings ($\g_L=\g_R$)}
    \label{fig:disc4}
\end{figure}

\begin{figure}
     [t]
     \center{\hspace{0cm}
         \epsfxsize = 5in
         \epsffile{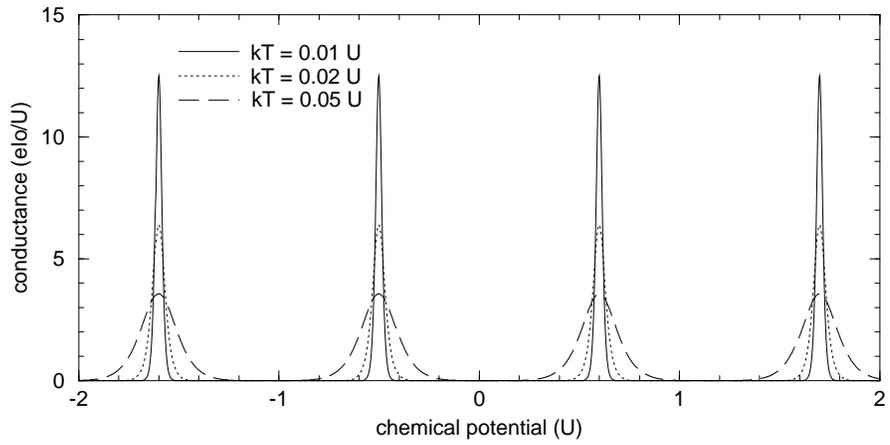}}
    \caption[Ohmic conductance through a dot with discrete levels]
            {Ohmic conductance through a dot with discrete levels
             ($\d_{\e} = 0.1 U$)}
    \label{fig:discond}
\end{figure}

\begin{figure}
     \center{\hspace{0cm}
         \epsfxsize = 5in
         \epsffile{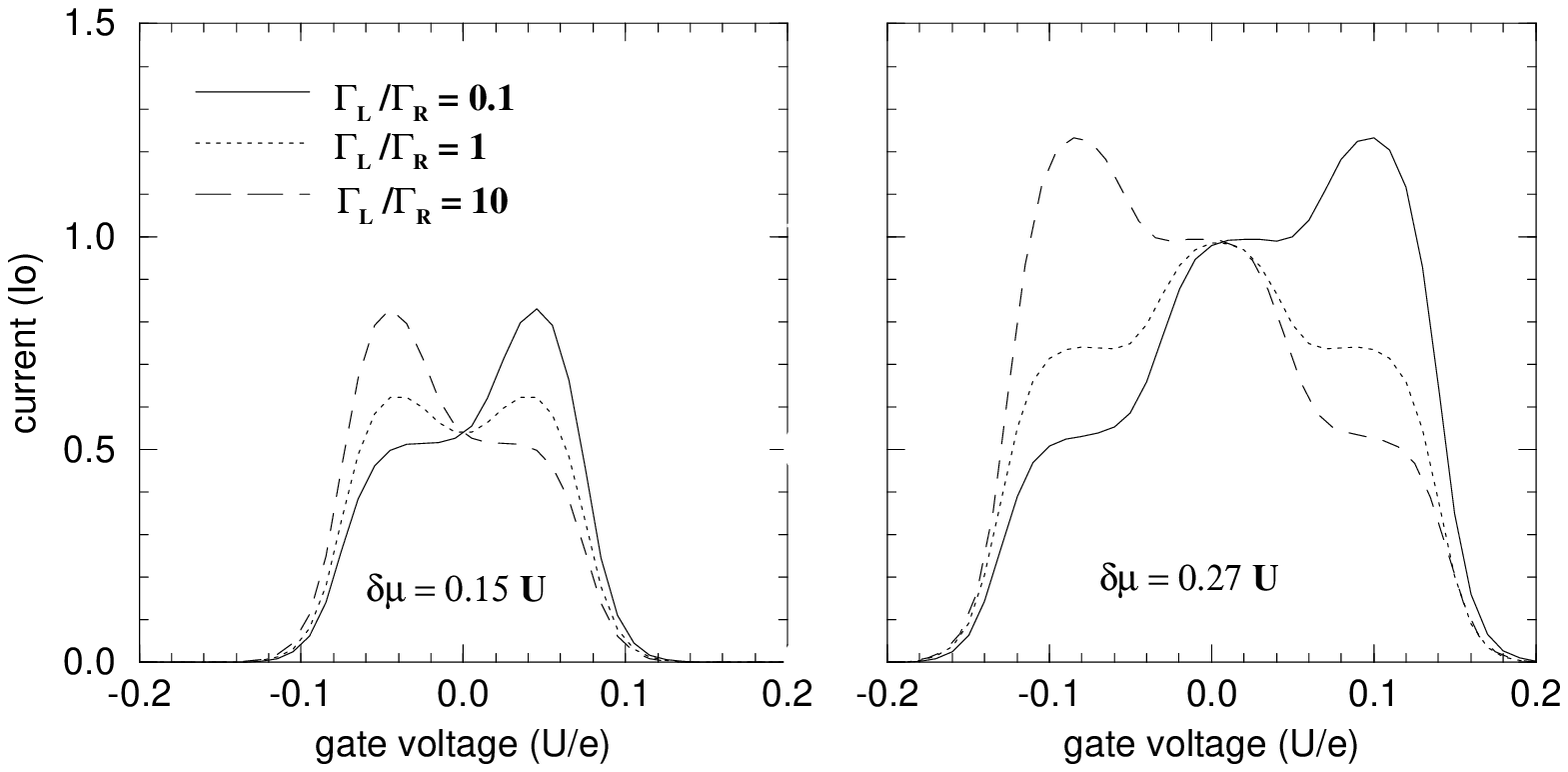}}
    \caption[Current {\it vs} gate voltage at fixed bias]
            {Current {\it vs} gate voltage at fixed bias ($\d_{\e} = 0.1 U$)}
    \label{fig:discgate}
\end{figure}

{
\noindent
When the energy level broadening is negligible compared to the temperature, 
i.e. in the limit of weak coupling to the reservoirs, the density of 
states in the dot can be adequately described by a set of delta functions.
The leads are assumed to be in thermal equilibrium, described by the 
Fermi-Dirac distributions $f_L$ and $f_R$. To a first approximation, 
the electron-electron interaction can 
satisfactorily be described using the charging model, where each pair of 
electrons has an associated Coulomb repulsion of $U$. 
Fermi's golden rule 
gives the tunnelling rates between the dot and the reservoirs for all 
energy levels $k$. 
\begin{eqnarray}
T_{L1}^k(N) &=& {\g_L \over \hbar} f_L(\e_k+N U) \\
T_{1L}^k(N) &=& {\g_L \over \hbar} [1-f_L(\e_k+N U)]
\end{eqnarray}
where $\g_L= 2 \pi \rho_L |V_L|^2$ is the strength of coupling to the left 
reservoir. The expressions for the tunnelling rates through the right barrier 
are similar. 
It is assumed that the quantum dot can be approximated by 
a parabolic confining potential so that the single particle energy levels
are equally spaced by an energy $\d_{\e}$. In figures \ref{fig:disc1} 
and \ref{fig:disc4} the current and its associated differential conductance
is plotted as a function of the applied bias for a range of energy level
spacings. 
In numerical calculations it is possible to take into account more 
realistic energy level spectra, enabling a closer comparison with 
experiment. When it is taken into account that the total spin of the 
system can only change by $1/2$ with each tunnelling event, it appears 
that negative differential conductance may occur in specific regions 
\cite{Weinmann1,Weinmann2}. 

In general a dot with closely packed energy levels yields a higher current 
than a dot with a sparse energy level spectrum, because of the higher number of 
current paths available. When $\d_{\e} \ll U$ then the metallic
regime is entered.

When the energy level spacing is not negligible, the I-V characteristics are
typified by two energy scales, the Coulomb repulsion energy $U$ and the
bare energy level spacing $\d_{\e}$. As in the metallic regime, one expects 
the I-V characteristic to display a current step whenever the maximum
occupation of the dot increases by one. This happens with a period $U+\d_{\e}$,
since an extra electron not only has to overcome the Coulomb barrier but also
has to tunnel to the next available energy level. In addition, there is 
also some fine structure which has an associated period of $\d_{\e}$. This is 
caused by the fact that an extra current path is created when the bias is 
increased by $\d_{\e}$. 

When $\g_L \gg \g_R$ the dot will be mostly maximally occupied and the most 
marked current increases occur whenever the Coulomb blockade can be overcome. 
This means that the period $U+\d_{\e}$ is accentuated. 
In the opposite regime $\g_L \ll \g_R$ the dominant period is the level 
spacing.
Under normal operating conditions $\g_L \simeq \g_R$ the two periods coexist
(see figure \ref{fig:disc4}). It is noted that at higher bias voltages the
number of peaks in the differential conductance increases, as new current
paths become available at different energies for different occupation numbers.
Only when the ratio $U/\d_{\e}$ is an integer do peaks corresponding to 
different occupation numbers coincide. 

The Ohmic conductance through a quantum dot (figure \ref{fig:discond}) differs
from that through a metallic dot in two
significant ways. Firstly, the periodicity of the conductance peaks has 
increased by the level spacing $\d_{\e}$. Secondly, the temperature 
dependence of the peaks has changed its nature. An increase in temperature
now not only leads to larger thermal broadening, but also to a lowering of 
the peak amplitude which is inversely proportional to the temperature. 
This is due to the fact that transport proceeds through 
a single energy level. The temperature dependence is proportional to the 
derivative of the Fermi-distribution function in the reservoirs \cite{Meirav}.
Therefore,
at low temperatures $k_BT \ll U$ the Ohmic conductance can be written as
\begin{equation}
G(\mu) = {e^2 \over \hbar} {\g_L \g_R \over \g_L + \g_R} \sum_N {1 \over 4 k_BT 
\cosh^2({\mu-\mu_N \over 2 k_BT})}
\end{equation}
where the successive charge degeneracy points $\mu_N$ are spaced by an energy 
$U+\d_{\e}$. For higher temperatures several levels should be taken into 
consideration, each of which is weighted by a factor determined by the 
Boltzmann distribution \cite{Been44}. 
\begin{equation}
G(\mu) = {e^2 \over \hbar} {\g_L \g_R \over \g_L+\g_R} \sum_N \sum_i
{1\over k_BT} P_i(N) f\left({\e_i + N U-\mu \over 2 k_BT}\right) 
\label{GBeen}
\end{equation}
where $P_i(N)$ is the joint probability that the dot contains $N$ electrons and 
that the single particle level $\e_i$ is empty. 
Note that the contributions from levels other than at positions
$\mu_N$ are too small to cause a peak in the conductance. 

Apart from the I-V characteristics and the Ohmic conductance, there is a 
third useful experiment that can be carried out. This involves applying a 
constant bias across the dot and studying the resultant current as a function
of the gate potential. Typically the source-drain bias is chosen to be 
less than the charging energy which isolates the effects of the 
energy separation of the zero dimensional states of the dot. In figure 
\ref{fig:discgate} the current has been calculated as a function of the 
gate voltage for a dot with a constant level spacing $\d_{\e} = 0.1 U$. The 
thermal energy is given by $k_BT = 0.01 U$.
In accordance with some recent experiments \cite{Johnson,Foxman,Vaart1} 
a number of peaks and troughs can be observed in the current. All peaks 
and troughs can be characterised by the number of levels available to an
incoming electron to tunnel onto, and the number of levels from which an
electron can tunnel out of the dot. For a Fermi level separation $\delta \mu$
of $0.15 U$ between the reservoirs (see figure \ref{fig:discgate}) 
these numbers are 
$1,2 \rightarrow 1,1 \rightarrow 2,1$. This explains why the total peak 
is split into two subpeaks. For the second set of graphs with $\delta \mu=
0.27 U$ the sequence of 
available tunnelling levels is $1,3 \rightarrow 1,2 \rightarrow 2,2 \rightarrow
2,1 \rightarrow 3,1$. It is also clear that asymmetric tunnelling barriers 
cause the current peaks to be asymmetric.

}

\section{Tunnelling through two dots in series} \label{quant_two}
\begin{figure}
     \center{\hspace{0cm}
         \epsfxsize = 5in
         \epsffile{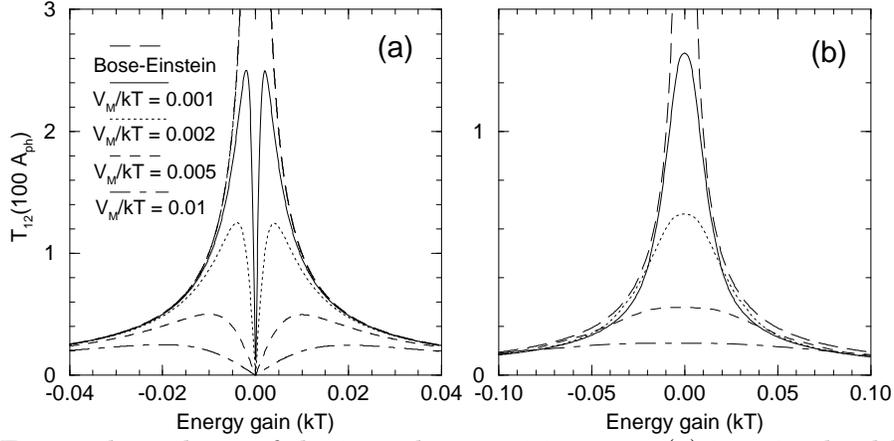}}
    \caption{Energy dependence of the inter-dot transition rate: 
    (a) ignoring level broadening, (b) including broadening $\g_{\phi} =
     5 V_M$.}
    \label{fig:T12form}
\end{figure}

{
\noindent
When studying the current and conductance properties of two quantum dots 
connected in series between the source and the drain, one obviously has 
to take into account the tunnelling between the dots. Inter-dot transitions 
are qualitatively different from transitions between a dot and a reservoir. 
The reservoir can be assumed to have a continuous density of states, so that
electrons can always tunnel elastically into and out of the reservoir. 
Even though inelastic
tunnelling events would in principle be allowed, their contribution would
be relatively small compared to the elastic tunnelling rate. 
Therefore inelastic scattering only has a significant effect on the
transport through a single dot if the scattering takes place inside the 
dot, i.e. relaxation. 

As far as tunnelling between dots is concerned, it is clear that the 
elastic tunnelling rate is significant only when the energy levels in the 
dots line up. This is obviously not generally the case. Usually an 
electron would have to interact inelastically in order to tunnel to a 
different energy level in the neighbouring dot. The energy difference 
would normally be absorbed or provided by phonons. 

As a theoretical model, one can consider the double dot system to be 
coupled to a phonon reservoir or a heat bath with a coupling strength 
$|V_{ph}|^2$. The energy spectrum of the independent oscillators of the 
phonon reservoir is characterised by the density of phonon states 
$\rho_{ph}(E)$. Some work has been done 
to calculate the current through a double-barrier resonant structure with 
some interaction between electrons and photons \cite{WinJac,Cai,Kouw}. 
The coupling to the optical phonons creates transmission subbands which are 
observable in the I-V characteristics.
In this section, where inter-dot transmission is considered, the interaction
with acoustic phonons will be the dominant mechanism. The energy spectrum 
for acoustic phonons is given by the Debye density of states.
\begin{equation}
\rho_{ph}(E) = C_{ph} \left\{ \begin{array}{ll}
                  E^2 & E< k_B \Theta_D \\
                  0   & E> k_B \Theta_D 
                  \end{array} \right.
\end{equation}
In a two-dimensional electron gas the Debye temperature $\Theta_D$ is 
approximately in the range $\Theta_D \sim 200-700 K$ \cite{Hess}. This 
is of the order of a few tens of $meV$, which corresponds to several 
times the Coulomb interaction energy $U$. This is typically larger than 
the voltage drop across the device, so that one can simply use the 
phonon density of states below the cut-off energy $k_B \Theta_D$.

Phonons are bosons, so that the Pauli principle does not apply 
and states can be multiply occupied. The average occupation number of 
states at a given energy is given by the Bose-Einstein distribution
$n_B(E)$.
\begin{equation}
n_B(E) = {\Theta(E) - \Theta(-E) \over e^{E/k_BT} -1}
\end{equation}
where $\Theta(E)$ is the Heaviside step function and $E$($-E$) is the 
energy provided (absorbed) by the phonon bath. The Bose-Einstein 
distribution at positive and negative values of $E$ differs by $1$, which
is indicative of the fact that a heat bath can always absorb energy.
 
A plausible model for the inelastic tunnelling rate 
$T_{12}^{kl}$ between an
energy level $\e_k$ in the first dot and $\e_l$ in the second dot is 
as follows
\begin{equation}
T_{12}^{kl} = {|V_{ph}|^2 \over \hbar} \int_{-\infty}^
{\infty} d\w \int_{-\infty}^{\infty} d\w' \rho_1^k(\w) \rho_2^l (\w') 
\left[\sum_{1,2} |\phi^k(\w) \phi^l(\w')|^2 \right] 
\rho_{ph}(\w'-\w) n_B(\w'-\w)
\label{trans}
\end{equation}
where $\rho_1^k(\w)$ and $\rho_2^l(\w')$ are the densities of state for the 
energy levels. Considering the dots in isolation from the reservoirs, the 
eigenstates cannot be regarded as localised in each of the dots, but the 
wavefunctions will leak slightly into the other dot by virtue of the 
hopping potential $V_M$. This results in the overlap integrals 
$|\phi_1^k(\w) \phi_1^l(\w')|^2$ and 
$|\phi_2^k(\w) \phi_2^l(\w')|^2$. Naively this can be
interpreted as the probability that the inelastic process can take place 
between the wavefunction components in the same dot. 
The wavefunctions are calculated from a
simple two-dimensional Hamiltonian with diagonal elements $\w$ and $\w'$ and 
off-diagonal elements given by the hopping potential $V_M$. 
By calculating its eigenvectors the 
overlap matrix can be determined. 
\begin{equation}
\sum_{1,2} |\phi^k(\w) \phi^l(\w')|^2 = {2 V_M^2 \over (\w'-\w)^2 + 4 V_M^2}
\end{equation}

\noindent
The density of states in the dots is broadened mainly as a result of the 
inelastic scattering, which is a prerequisite for the incoherent regime. 
To a first approximation the broadening can be considered as a Lorentzian
distribution \cite{StoneLee} with a broadening $\g_{\phi}$  
proportional to the phase-breaking rate. Now equation \ref{trans} can 
be rewritten as a single integral over the energy difference $E=\w'-\w$ 
between the initial and final state.
\begin{equation}
T_{12}^{kl} = A_{ph}
\int_{-\infty}^{\infty} dE {\g_{\phi}/\pi \over (E+\e_k-\e_l)^2 + \g_{\phi}^2}
{E^2 \over E^2 + 4 V_M^2}  n_B(E)
\end{equation}
with $A_{ph} = 2 V_M^2 |V_{ph}|^2 C_{ph}/\hbar$. 

It must be stressed that the broadening of the energy levels should be 
much less than the temperature. Otherwise it is not valid anymore to 
use the Boltzmann distribution to calculate the occupation probabilities.
This assumption is consistent with the calculation of the reservoir-dot 
transition 
rate of section \ref{quant_one}. Figure \ref{fig:T12form}a shows the 
energy dependence of the inter-dot tunnelling rate for the case 
when the energy-levels are considered 
to be $\delta$-functions. However, it should still be taken into account 
that the level broadening is large compared to the coupling between the
dots, causing the peaks in the transition rate at $E=\pm 2 V_M$ to be 
smeared. At energy differences $\d=\e_l-\e_k$ which are small 
compared to the broadening, the transition rate can now be approximated by 
a Lorentzian  of width $2 \g_{\phi}/\hbar$ and height 
$2 A_{ph} {\rm ln}[\g_{\phi}/2 V_M]
/\pi \g_{\phi}$ (see figure \ref{fig:T12form}b).
At larger energy differences $|\d| > \Gamma $ the 
function is sufficiently close to the Bose-Einstein distribution. 

As the broadening is much smaller than the thermal energy, the inter-dot
transition rate for small $\d$ 
can get much larger than the transition rate between dot 
and reservoir so that the inter-dot transition is no longer the 
current-limiting process. In this case it is of no relevance whether the 
transition rate between the dots is infinite or simply very large. 
This suggests that the inter-dot transition rate may be approximated by 
the Bose-Einstein distribution $A_{ph} n_B(\d)$ at all values of $\d$. 
This makes the mathematical analysis much more transparent.

When one calculates the transition rate between dots with a given occupation
number, one should sum over all possible tunnelling events according to 
equation \ref{Tgen}. The greatest contribution should come from the energy 
range where the electron can tunnel from levels which are mostly 
occupied to levels which are mostly empty. Outside this window 
the product $P_1(k|N_1+1)[1-P_2(l|N_2)]$ falls off exponentially.
However, since the Bose-Einstein distribution diverges at $\d \rightarrow
0$, the total transition rate seems to be dominated by matching levels,
even when they are situated at energies far removed from the Fermi level. 
This is clearly an unphysical situation. This anomaly can be removed 
by again including the broadening in the calculation of the transition rate.
Alternatively, it can be argued that the transition rate for a given pair 
of levels is the combination of the previously defined rate and the 
rate at which electrons can get into an excited level. This second rate is the 
relaxation rate $\hbar/\tau_{\rm rel}$. Remembering that the inverses of 
the rates of consecutive processes add together, the total inter-dot rate 
can be obtained for given occupation numbers.  
\begin{equation}
T_{12}(N_1,N_2) = \sum_{k,l} P_i(k|N_i+1)[1-P_{i\pm1}(l|N_{i\pm1})]
  {T_{12}^{kl} \hbar/\tau_{\rm rel} \over T_{12}^{kl} + \hbar/\tau_{\rm rel}}
\end{equation}

\noindent
In order to perform some realistic simulations it is helpful 
to know the dependence on the size of both the charging energy $U$ and the 
confinement energy $\d_{\e}$. In the charging model approximation the 
Coulomb repulsion is inversely proportional to the capacitance $C$ of the dot.
This implies that the repulsion energy $U$ is also inversely proportional to 
the area of the dot. The single particle energy level spacing is given by
\cite{Staring}
\begin{equation}
\d_{\e} = {\pi \hbar^2 \over m^* {\rm area}}
\end{equation}
where $m^*$ is the effective mass of an electron in the two dimensional 
electron gas in which the quantum dot has been defined. It follows that
both the Coulomb energy and the level spacing scale inversely proportionally
to the area of the quantum dot. 
}

\subsection{Ohmic conductance}
\begin{figure}
     \center{\hspace{0cm}
         \epsfxsize = 5in
         \epsffile{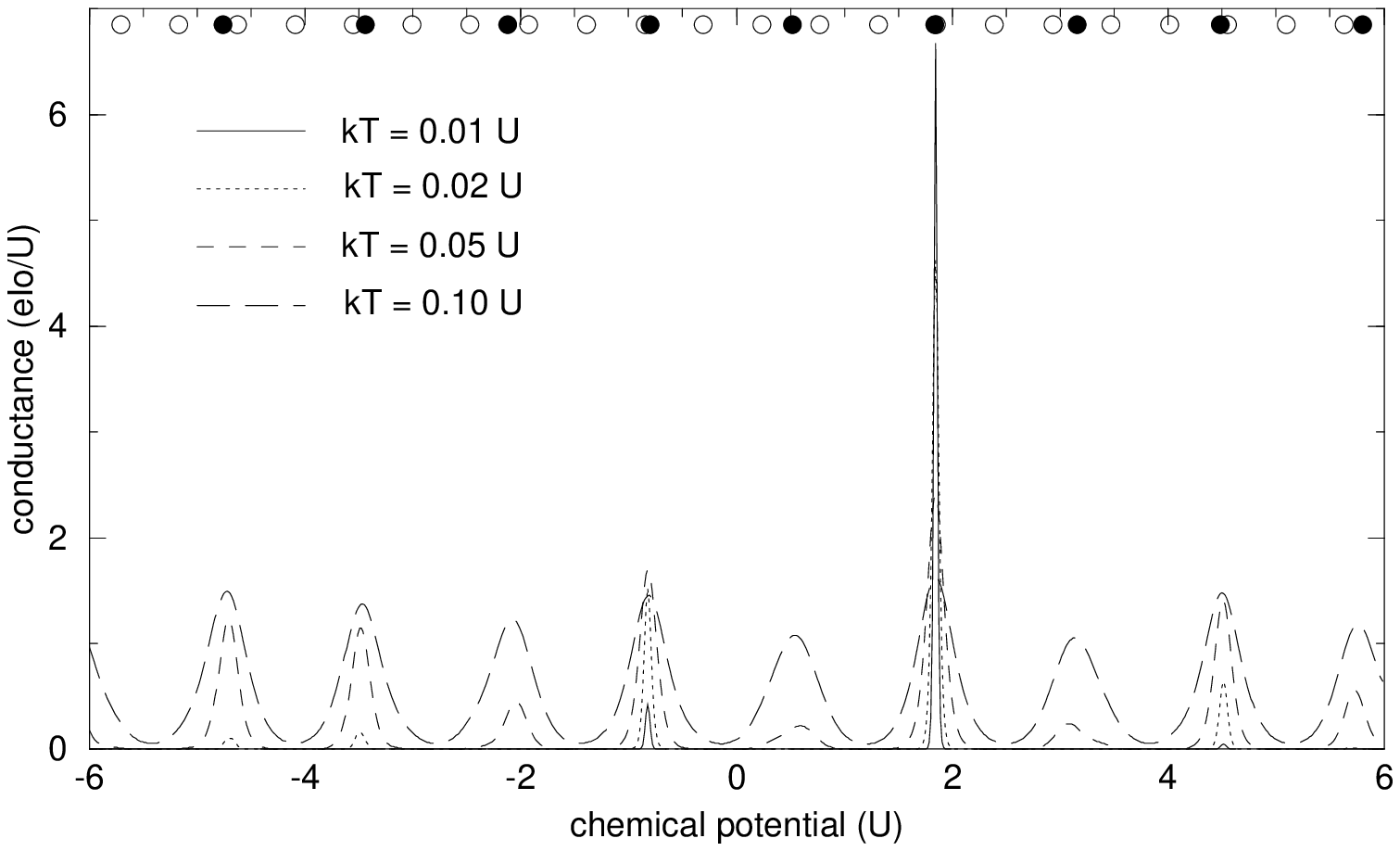}}
    \caption{Ohmic conductance through a double dot with 
             $A_{ph}\hbar/\Gamma = 1$ ($U_1=0.41 U$, $U_2=U$, $\d_{\e_1}=0.13 U$,
             $\d_{\e_2} = 0.32 U$). The empty and filled circles indicate the 
             positions at which the average occupation increases by one 
             for dot $1$ and $2$ respectively.}
    \label{fig:cond2}
\end{figure}

\begin{figure}
     \center{\hspace{0cm}
         \epsfxsize = 3in
         \epsffile{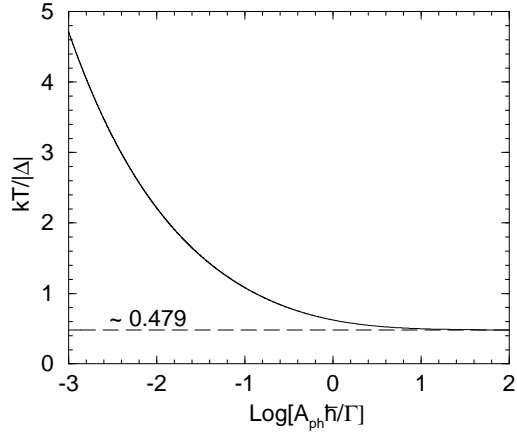}}
    \caption{Temperature at which a conductance peak reaches its maximum}
    \label{fig:max}
\end{figure}

\begin{figure}
     \center{\hspace{0cm}
         \epsfxsize = 5in
         \epsffile{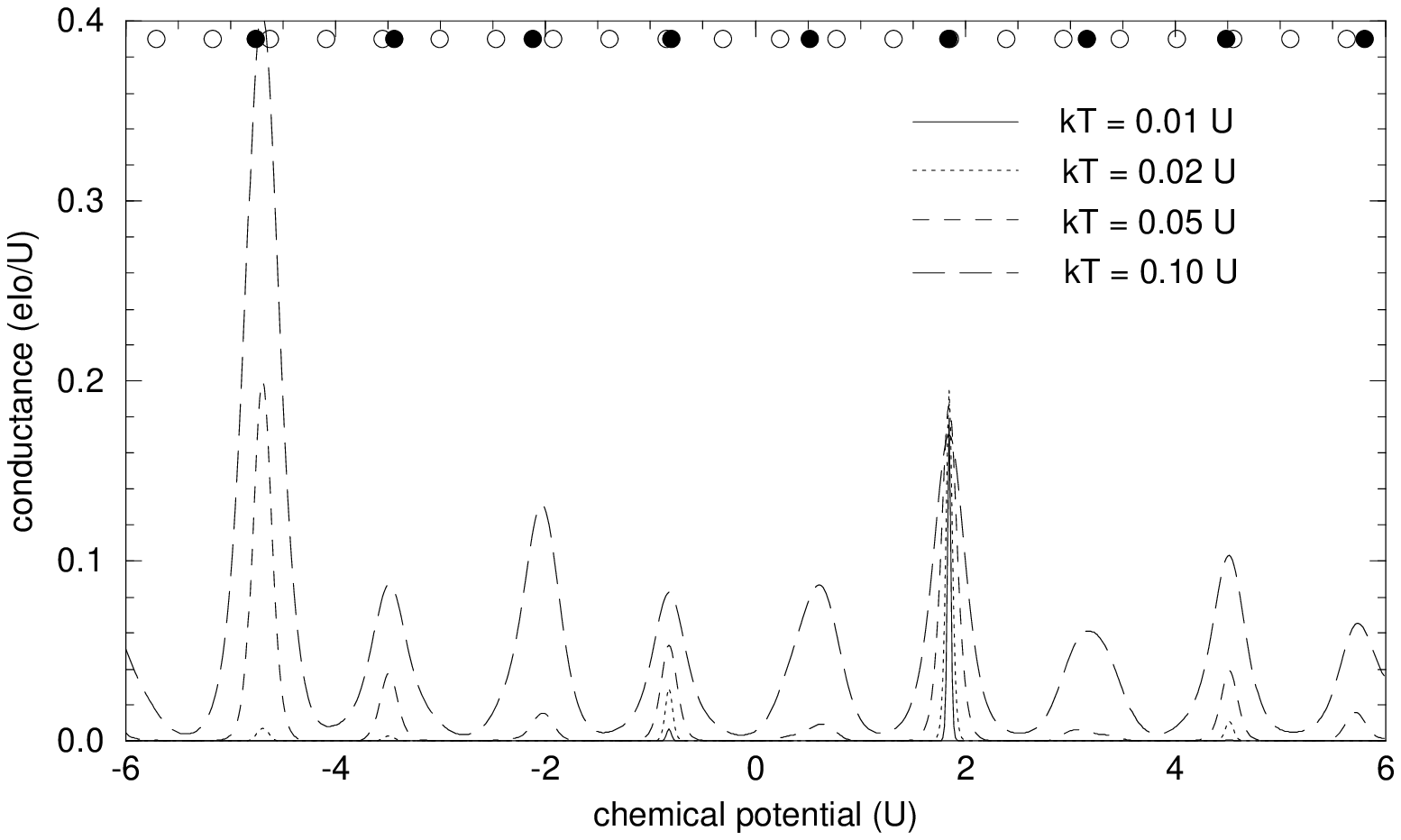}}
    \caption{Ohmic conductance through a double dot with 
             $A_{ph}\hbar/\Gamma = 0.01$ ($U_1=0.41 U$, $U_2=U$, 
             $\d_{\e_1}=0.13 U$, $\d_{\e_2} = 0.32 U$). 
             The empty and filled circles indicate the
             positions at which the average occupation increases by one
             for dot $1$ and $2$ respectively.}
    \label{fig:cond4}
\end{figure}

{
\noindent
Figure \ref{fig:cond2} shows the Ohmic conductance through two dots with 
a relatively strong inelastic tunnelling coefficient $A_{ph}$.
In order to analyse the structure of the peaks, assume that the total 
occupancy of each dot can only fluctuate by one electron. Moreover, assume that 
only a single level per dot (at the charge degeneracy points) contributes 
substantially to the transport. 
Then the global master equation 
is used to obtain an expression for the Ohmic conductance.
\begin{eqnarray}
G &=& {e^2 A_{ph} \g_L \g_R (\g_L+\g_R) \over
    4 k_BT \cosh\left({\mu-\e_k \over 2 k_BT}\right) 
         \cosh\left({\mu-\e_l \over 2 k_BT}\right)} \times
 \left[ A_{ph} \hbar 
   \left( {\g_L^2 \cosh\left({\mu-\e_l \over 2 k_BT}\right) \over 
        \cosh\left({\mu-\e_k \over 2 k_BT}\right)} + 
          {\g_R^2 \cosh\left({\mu-\e_k \over 2 k_BT}\right) \over 
        \cosh\left({\mu-\e_l \over 2 k_BT}\right)} \right)
  \right.
  \nonumber \\ 
&+& \left. 2 A_{ph}\hbar\g_L\g_R \cosh\left({\e_k-\e_l \over 2 k_BT}\right)
 + 2 \g_L \g_R (\g_L+\g_R) \sinh\left({|\e_k-\e_l| \over 2 k_BT}\right)
\right]^{-1}
\end{eqnarray}
When the two reservoirs are equally strongly coupled to the dots, i.e. 
$\g_L = \g_R = \g$, then the above expression peaks at $\mu=(\e_k+
\e_l)/2$. The peak conductance is given by (using $\d = 
\e_l-\e_k$)
\begin{equation}
G_{\rm max} = {e^2 A_{ph} \g \over 
     8 k_BT \cosh^3 \left({\d \over 4 k_BT}\right) 
     \left[ A_{ph} \hbar \cosh\left({\d \over 4 k_BT}\right) +
       2 \g \sinh\left({|\d| \over 4 k_BT}\right) \right] }
\end{equation}
When the peak height is investigated as a function of the 
temperature, then it appears that it has a maximum at a value of 
$k_BT/|\d|$ which is given by the solution of the following 
transcendental equation.
\begin{equation}
\left(2 {A_{ph} \hbar \over \g} 
{|\d|\over k_BT} + 3 {|\d|\over k_BT} - 4 \right) 
\tanh\left({|\d| \over 4 k_BT}\right) = 
2  {A_{ph}\over \g} - {|\d|\over k_BT} 
\end{equation}

\noindent The solution is plotted in figure \ref{fig:max}. When the ratio
of the inelastic tunnelling coefficient to the reservoir coupling 
$A_{ph}\hbar/\Gamma$ is of order unity or larger, then the temperature at 
which a particular conductance peak reaches its maximum height is given by
roughly half the energy difference $|\d|$. This maximum height is 
larger as the energy difference $\d$ gets smaller. 
This calculation is in good quantitative agreement with  
the temperature dependence of the 
conductance curves of figure \ref{fig:cond2}. At higher temperatures 
the calculation becomes more inaccurate as several levels and more 
occupation numbers will start to contribute to the transport. 

In figure \ref{fig:cond4} the conductance is shown for a double dot system 
which is identical to that of figure \ref{fig:cond2} with the exception that
the inelastic tunnelling coefficient is two orders of magnitude smaller. 
From figure \ref{fig:max} one would expect the peaks to be maximised at 
a temperature $k_BT \simeq 2.2 |\d|$. The above description of the 
conductance peaks seems to apply to most of the peaks. However, it is clear 
that the peak which is situated at $\mu \simeq -4.7 U$ has an anomalous 
behaviour. The conductance at this point is much larger than expected. 
This is due to the fact that one of the lower levels in the first dot 
very nearly matches up with the dominant level in the second dot, thus 
strongly enhancing the inter-dot tunnelling rate. 

This effect most strongly shows up when the inelastic tunnelling coefficient
$A_{ph} \hbar$ is small compared to the coupling $\g$ to the reservoirs. 
This is simply due to the fact that the inter-dot tunnelling is the main 
current-limiting process and will therefore tend to dominate the physics. 
For large inelastic tunnelling coefficients the current will mainly be 
determined by the matching of the levels between dot and reservoir. This
explains why the afore-mentioned effect is almost imperceptible in figure
\ref{fig:cond2}. 
}

\subsection{Current characteristics}
\label{quant_2_curr}
\begin{figure}
     \center{\hspace{0cm}
         \epsfxsize = 5in
         \epsffile{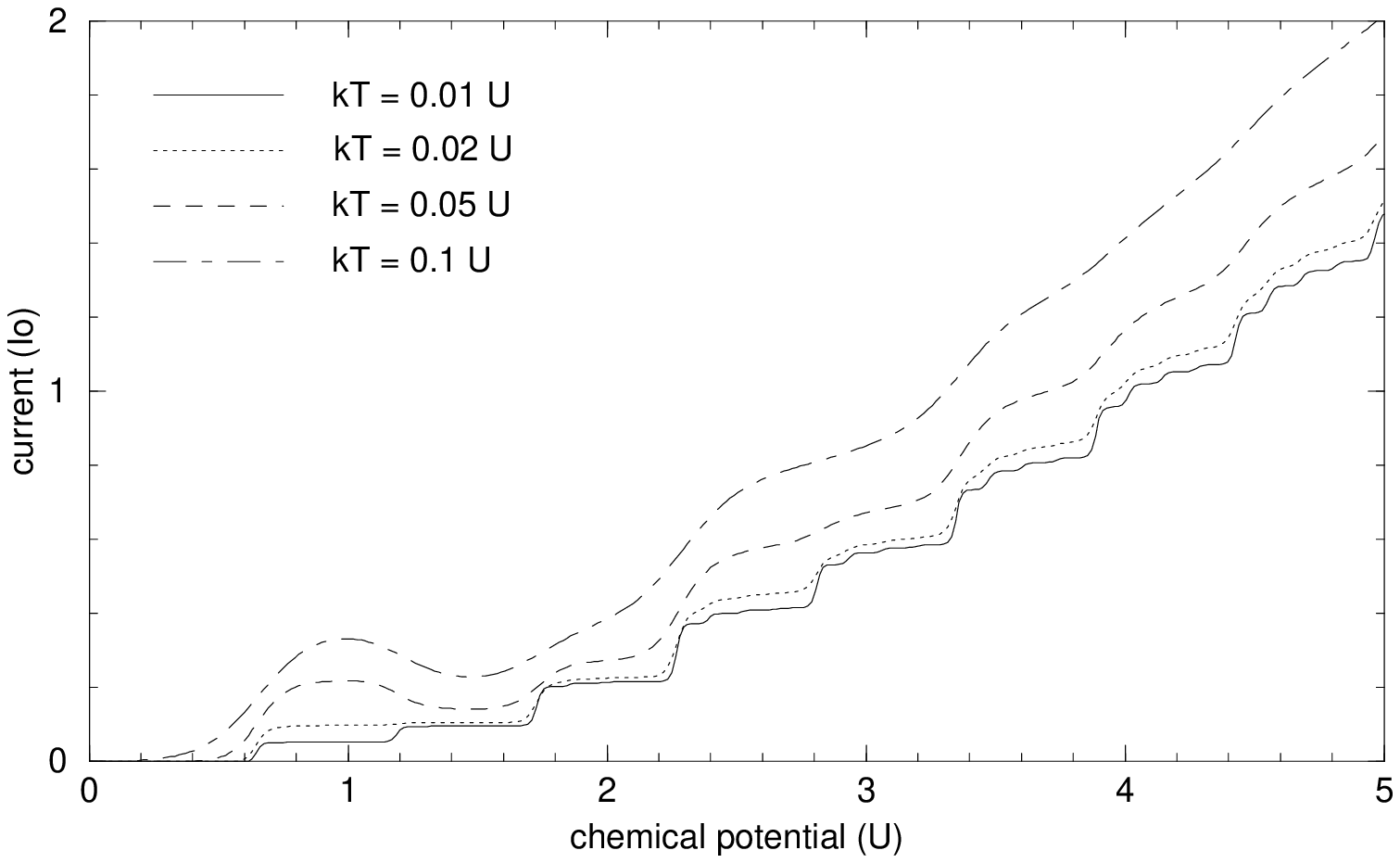}}
    \caption[I-V characteristics for a double dot at various temperatures]
            {I-V characteristics for a double dot at various temperatures
            ($U_1=0.41 U$, $U_2=U$, $\d_{\e_1}= 0.13 U$, $\d_{\e_2}= 0.32 U$, 
             $A_{ph}\hbar/\g = 0.01$)}
    \label{fig:ycurr4}
\end{figure}

\begin{figure}
     \center{\hspace{0cm}
         \epsfxsize = 5in
         \epsffile{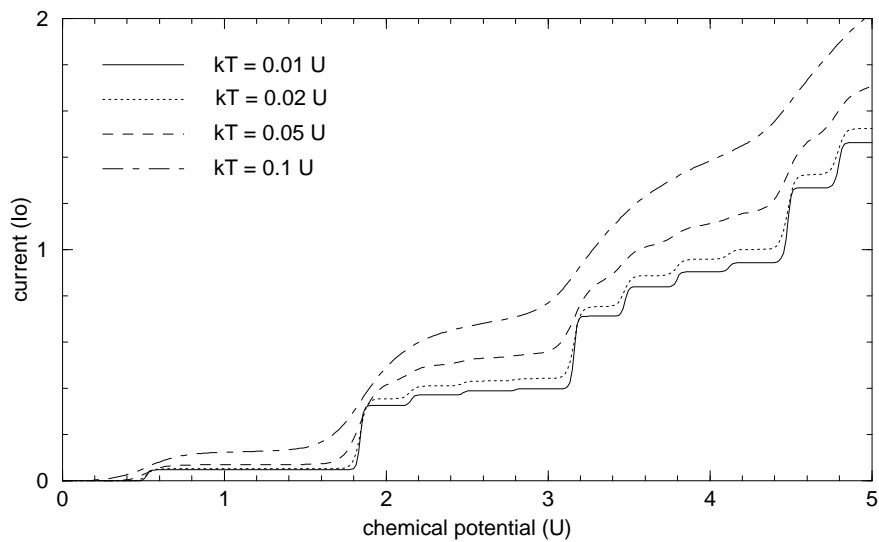}}
    \caption{I-V characteristics for a double dot at various temperatures
         (the dots are interchanged with respect to the previous figure)}
    \label{fig:yrcurr4}
\end{figure}

\begin{figure}
     \center{\hspace{0cm}
         \epsfxsize = 4in
         \epsffile{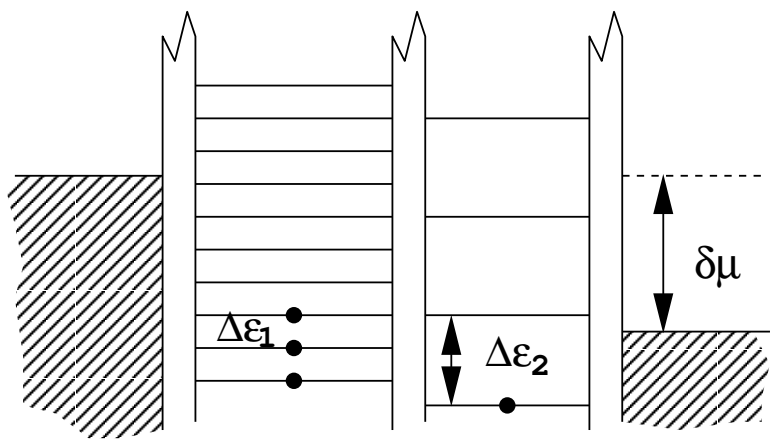}}
    \caption{Energy diagram for tunnelling through two dots at fixed bias}
    \label{fig:gate2}
\end{figure}

\begin{figure}
     \center{\hspace{0cm}
         \epsfxsize = 5in
         \epsffile{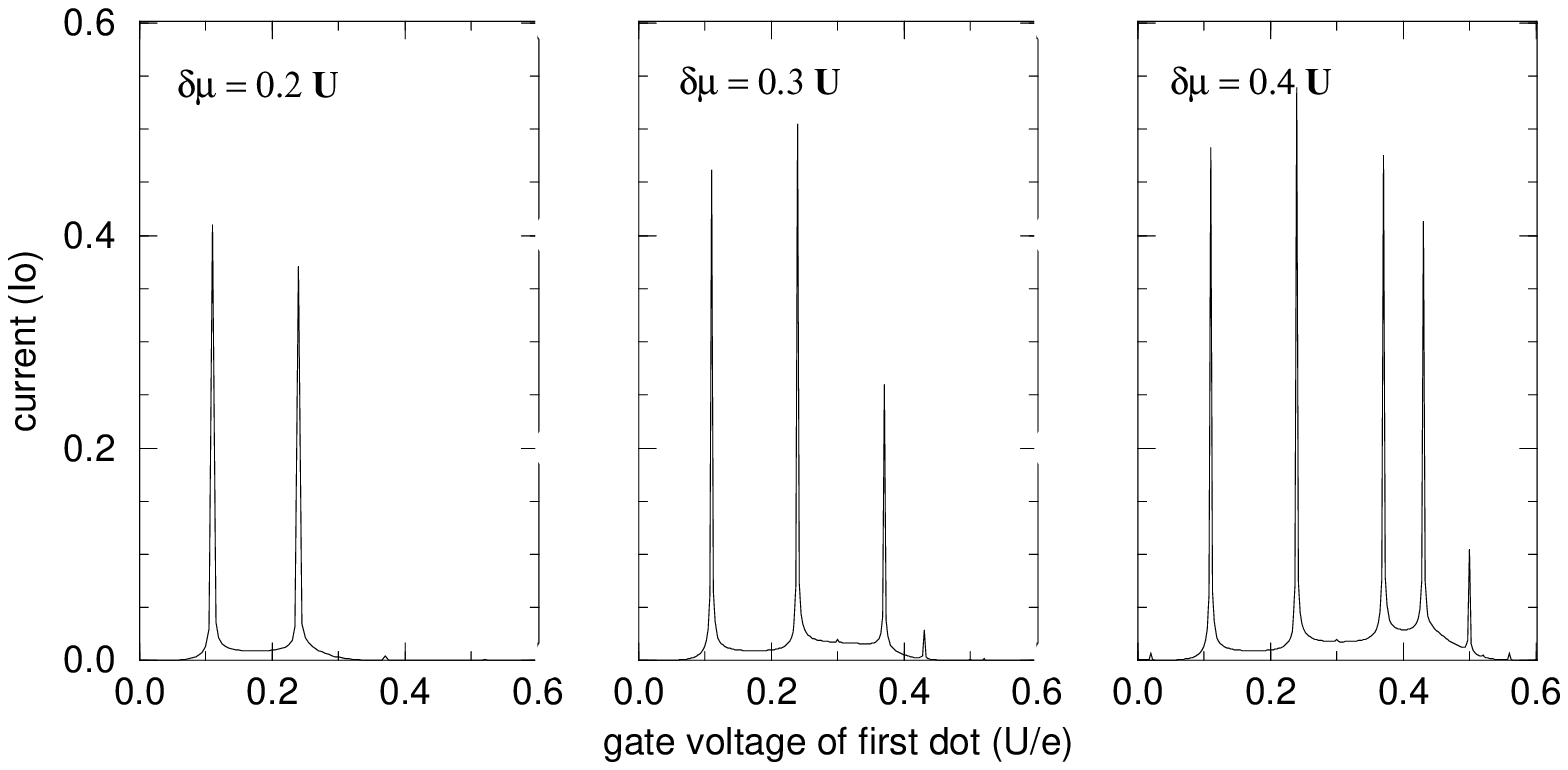}}
    \caption[Current {\it vs} gate voltage of first dot at fixed bias]
             {Current {\it vs} gate voltage of first dot at fixed bias
             ($U_1 = 0.41 U$, $U_2 = U$, 
             $\d_{\e_1} = 0.13 U$, $\d_{\e_2} = 0.32 U$, $k_BT = 0.02 U$, 
             $\g_L=\g_R=\g$, $A_{\rm ph}\hbar/\g = 0.01$)}
    \label{fig:gate2d}
\end{figure}

{
\noindent
The current-voltage characteristics have been calculated 
in figure \ref{fig:ycurr4} for a double dot 
system with specifications as indicated in the caption. 
Figure \ref{fig:yrcurr4} shows the results for 
a system where the dot specifications have been swapped. Physically this 
simply amounts to varying the chemical potential in the right reservoir 
instead of the left reservoir. 

At low temperatures two periods may be observed in the I-V characteristics.
The larger period is given by the Coulomb repulsion in the first dot plus a 
single 
particle spacing $U_1+\d_{\e_1}$. The smaller period is simply given by 
the single particle spacing $\d_{\e_1}$. This behaviour is reminiscent of 
the current through a single dot with $\g_L \gg \g_R$ (see section 
\ref{quant_one}).
In other words, the second dot with its surrounding barriers acts as a 
single barrier with a reduced transparency. The electrons which have entered 
the first dot will tunnel into a lower energy level in the second dot at a 
rate of approximately $A_{ph}$ since the levels in the dots will not 
normally line up. At larger temperatures the current curves will lose some 
features due to thermal smearing. This is clearly the case in figure 
\ref{fig:yrcurr4}. 

However, the current curve of figure \ref{fig:ycurr4} has a more 
complicated form at high temperatures. It has a region of negative 
differential conductance which occurs in the bias range where the average 
occupancy of the first dot increases from one to two electrons with 
respect to the average occupation number at zero bias. This can be 
explained by considering the (unlikely) case where a pair of energy levels 
matches up for a given set of occupation numbers. 
This will result in a highly increased tunnelling rate between 
the dots. When the bias across the device is raised the average occupancy 
of the dot will increase. This means that the dots are less likely to 
contain the number of electrons for which the levels lined up. 
Consequently the tunnelling rate between the dots and hence the overall 
current will decrease, in spite of the fact that the total number of 
tunnelling paths is likely to have increased. 

The temperature at which negative differential conductance can occur 
(if at all) is set by the energy difference of the matching pair of 
levels in the dots. Their energy separation has to be significantly less 
than $k_BT$ for the tunnelling rate to increase dramatically. A higher 
temperature can cause some levels to match up which could not really be 
considered energetically close at lower temperatures. However, a higher 
temperature also means that there is a smaller probability that 
the total inter-dot tunnelling is dominated by just a single pair of 
levels. This will reduce the effect. The conclusion is that negative 
differential conductance is more likely to occur at higher temperatures, but 
when it occurs at a lower temperature the effect will be more pronounced. 

Similar to the one dot case, one can extract information about the level 
spacing when one considers the current through the system at fixed bias 
$\delta\mu$ while varying 
the gate voltage of one of the dots, a measurement first performed by 
van der Vaart {\it et al.} \cite{VaartGodijn}.
This is depicted in 
figure \ref{fig:gate2}. The Coulomb repulsion energies and the level 
spacings are the same as before. The  inelastic tunnelling coefficient 
is small compared to the dot-reservoir coupling. 
The empty levels are the levels that an excess 
electron can occupy. Note that no more than a single extra electron can 
be contained in the dot, since this is prevented by the Coulomb blockade. 
Since electrons carry a negative charge, a rise in the gate potential 
causes a downward shift of the energy levels.

The results are shown in figure \ref{fig:gate2d} for a range of values of 
$\delta \mu$. A few generic features can be noted which are also true 
for tunnelling through a single dot.
Firstly, the current is periodic in the gate voltage with a period 
$U_1+\d_{\e_1}$. Only a single period is shown in the figure. 
Secondly, a significant current is only allowed to flow when the 
Coulomb blockade is overcome in both dots. This requires the lowest 
available level in each dot to lie within the energy window $\delta \mu$. 
This is clearly shown in figure \ref{fig:gate2d} where a larger bias voltage 
allows the current to flow at more values of the gate potential. 

The most striking feature of figure \ref{fig:gate2d} is the occurrence of 
sharp current peaks. These happen at values of the gate voltage where one 
of the occupied states of the first dot (containing the excess electron)
lines up with an empty level in the second dot. It is clear that this 
will happen with a periodicity $\d_{\e_1}$ (in the figure $\d_{\e_1} = 
0.13 U$). When the bias voltage is 
such that several levels in the second dot are contained within the 
energy window $\delta \mu$, then peaks will also occur at intervals 
$\d_{\e_2}$. This is the case in the third graph of figure \ref{fig:gate2d},
where $\d_{\e_2} = 0.32 U$. 

Finally, note that the current in the valleys between the peaks increases 
more or less linearly with the number of peaks.
This reflects the fact that at a higher gate potential there are
more electrons in the first dot which are able to tunnel downwards in energy
into the second dot. Each tunnelling path has an associated 
off-resonance tunnelling rate of $A_{ph}$ which will contribute an 
amount $A_{ph}\hbar I_0/\g$ towards the total current (assuming 
$A_{ph}\hbar \ll \g$). 
Therefore the current minima should increase by this amount every time 
the gate potential is increased by an amount $\d_{\e_1}$.

This seems to be a much more powerful method for determining the single particle 
level spacing than the analogous experiment with a single dot. 
In the model used in this chapter, the 
line shape will be a multiple of the Bose-Einstein distribution. However, 
at very small energy separations $\d$ the intrinsic width of the levels becomes 
significant
and the lineshape will be approximately a Lorentzian of width $2 \g_{\phi}$ near 
resonance but will be asymmetric off-resonance. This seems to 
corresponds at least qualitatively to recent experiment \cite{VaartGodijn}.
In the case of large $A_{\rm ph}\hbar/\g$, the middle barrier is no longer the 
current limiting barrier and the current will not peak anymore but simply 
increase with $\d$ until a saturation value of ${2 \over 3} e^2 /\hbar$ is
reached. 
Further experiments will have 
to show whether it is justified to assume that the inelastic tunnelling rate
can be approximated by the Bose-Einstein distribution. If this turns out 
to be a false assumption, then the master equation can still be used to model
the effect of a more realistic inelastic tunnelling rate. 

}

\section{Conclusions}
{
\noindent
For a dot with discrete levels, the distribution function at equilibrium 
differs from the Fermi-Dirac distribution, especially at low temperatures. 
The peaks in the Ohmic conductance are spaced by an amount $U+\d_e$.
In the I-V characteristics the effect of both the charge quantisation and 
the size quantisation can be observed. 

The transport through a double dot has been investigated assuming that 
inelastic transport through the inter-dot barrier can take place by means
of interaction with acoustic phonons. The main peaks in the Ohmic 
conductance reach a maximum at a temperature $k_BT$ which is at least roughly 
half the energy difference $\d$ between the dominant levels in the two dots. 
Other levels in the two dots that are well aligned can also have a 
significant effect on the conductance, especially when the inter-dot coupling
is weak. The I-V characteristics may contain regions of negative 
differential conductance. This is less likely to occur at low temperatures, 
although its effect will be stronger than at higher temperatures. 

The inter-dot spacing can be analysed spectroscopically by investigating the 
current through the double dot at fixed bias as a function of one of the 
gate voltages. This produces very narrow peaks with a width that is closely
related to the intrinsic level width. This method eliminates 
thermal smearing of the peaks.
}

\begin{references}

 \newcommand{\ph}   [1]{{\it Physica}                 {\bf #1}}
 \newcommand{\phb}  [1]{{\it Physica B}               {\bf #1}}
 \newcommand{\pr}   [1]{{\it Phys.\ Rev.}             {\bf #1}}
 \newcommand{\prbe}  [1]{{\it Phys.\ Rev.\ B}          {\bf #1}}
 \newcommand{\prle} [1]{{\it Phys.\ Rev.\ Lett.}      {\bf #1}}
 \newcommand{\zpb}  [1]{{\it Z.\ Phys.\ B}            {\bf #1}}
 \newcommand{\jpc}  [1]{{\it J.\ Phys.\ C}            {\bf #1}}
 \newcommand{\jpf}  [1]{{\it J.\ Phys.\ F}            {\bf #1}}
 \newcommand{\jpcm} [1]{{\it J.\ Phys.:\ Cond.\ Mat.} {\bf #1}}
 \newcommand{\jetp} [1]{{\it Sov.\ Phys.\ JETP}       {\bf #1}}
 \newcommand{\jetpl}[1]{{\it Sov.\ Phys.\ JETP\ Lett.}{\bf #1}}
 \newcommand{\revmp}[1]{{\it Rev.\ Mod.\ Phys.}       {\bf #1}}
 \newcommand{\ibm}  [1]{{\it IBM\ J.\ Res.\ Dev.}     {\bf #1}}
 \newcommand{\surf} [1]{{\it Surf.\ Sc.}              {\bf #1}}
 \newcommand{\pla}  [1]{{\it Phys.\ Lett.\ A}         {\bf #1}}
 \newcommand{\jltp} [1]{{\it J.\ Low\ Temp.\ Phys.}   {\bf #1}}
 \newcommand{\jappl}[1]{{\it J. Appl.\ Phys.}         {\bf #1}}
 \newcommand{\appl} [1]{{\it Appl.\ Phys.\ Lett.}     {\bf #1}}
 \newcommand{\eupl} [1]{{\it Europhysics\ Lett.}      {\bf #1}}
 \newcommand{\reppp}[1]{{\it Rep. Prog. Phys.}        {\bf #1}}
 \newcommand{\jpsj} [1]{{\it J. Phys. Soc. Japan}     {\bf #1}}
 \newcommand{\ssc}  [1]{{\it Solid State Comm.}       {\bf #1}}
 \newcommand{\ssp}  [1]{{\it Solid State Phys.}       {\bf #1}}
 \newcommand{\jmp}  [1]{{\it J. Math. Phys.}          {\bf #1}}
 \newcommand{\advp} [1]{{\it Adv. Phys.}              {\bf #1}}
 \newcommand{\phm}  [1]{{\it Philos. Mag.}            {\bf #1}}


\bibitem{Wees} B.J. van Wees, H. van Houten, C.W.J. Beenakker, J.G. Williamson,
L.P. Kouwenhoven, D. van der Marel, C.T. Foxon, \prl{60}, 848 (1988).

\bibitem{Wharam} D.A. Wharam, T.J. Thornton, R. Newbury, M. Pepper, H. Ahmed,
J.E.F. Frost, D.G. Hasko, D.C. Peacock, D.A. Ritchie, G.A.C Jones,
\jpc{21}, L209 (1988).

\bibitem{Scott-Thomas} J.H.F. Scott-Thomas, S.B. Field, M.A. Kastner,
H.I. Smith, D.A. Antoniadis, \prl {62}, 583 (1989).

\bibitem{HoutBeen} H. van Houten, C.W.J. Beenakker, \prl {63}, 1893 (1989).

\bibitem{Reed} M.A. Reed, J.N. Randall, R.J. Aggarwal, R.J. Matyi, T.M. Moore,
A.E. Wetsel, \prl{60}, 535 (1988).

\bibitem{Amman2} M. Amman, R. Wilkins, E. Ben-Jacob, P.D. Maker, R.C Jaklevic,
\prb{43}, 1146 (1991).

\bibitem{Ruzin} I.M. Ruzin, V. Chandrasekhar, E.I. Levin, L.I. Glazman, 
\prb {45}, 13469 (1992).

\bibitem{Kemerink} M. Kemerink, L.W. Molenkamp, \appl{65}, 1012 (1994).

\bibitem{Lee} Y. Meir, N.S. Wingreen, P.A. Lee, \prl {66}, 3048 (1991).

\bibitem{Been44} C.W.J. Beenakker, \prb{44}, 1646 (1991). 

\bibitem{Averin2} D.V. Averin, A.N. Korotkov, K.K. Likharev, \prb{44}, 6199 
(1991).

\bibitem{Weinmann1} D. Weinmann, W. H\"{a}usler, W. Pfaff, B. Kramer, 
U. Weiss, \eupl{26}, 467 (1994).

\bibitem{Weinmann2} D. Weinmann, W. H\"{a}usler, B. Kramer, \prl{74}, 984
(1995).

\bibitem{Meirav} U. Meirav, M.A. Kastner, S.J. Wind, \prl {65}, 771 (1991).

\bibitem{Johnson} A.T. Johnson, L.P. Kouwenhoven, W. de Jong, N.C. van der 
Vaart, C.J.P.M. Harmans, C.T. Foxon, \prl{69}, 1592 (1992).

\bibitem{Foxman} E.B. Foxman, P.L. McEuen, U. Meirav, N.S. Wingreen, Y. Meir,
P.A. Belk, N.R. Belk, M.A. Kastner, S.J. Wind, \prb{47}, 10020 (1993).

\bibitem{Vaart1} N.C. van der Vaart, A.T. Johnson, L.P. Kouwenhoven, 
D.J. Maas, W. de Jong, M.P. de Ruyter van Steveninck, A. van der Enden,
C.J.P.M. Harmans, \phb{189}, 99 (1993). 

\bibitem{WinJac} N.S. Wingreen, K.W. Jacobsen, J.W. Wilkins, \prl{61},1396
(1988).

\bibitem{Cai} W. Cai, T.F. Zheng, P. Hu, B. Yudanin, M. Lax, 
\prl{63}, 418 (1989).

\bibitem{Kouw} L.P. Kouwenhoven, S. Jauhar, K. McCormick, D. Dixon, P.L. McEuen,
Y.V. Nazarov, N.C. van der Vaart, C.T. Foxon, \prb{50}, 2019 (1994).

\bibitem{Hess} K. Hess, C.T. Sah, \prb{10}, 3375 (1974). 

\bibitem{StoneLee} A.D. Stone, P.A. Lee, \prl{54}, 1196 (1985). 

\bibitem{Staring} A.A.M. Staring, H. van Houten, C.W.J. Beenakker,
\prb {45}, 9222 (1992).

\bibitem{VaartGodijn} N.C. van der Vaart, S.F. Godijn, Y.V. Nazarov, 
C.J.P.M. Harmans, J.E. Mooij, \prl{74}, 4702 (1995). 

\end{references}
\end{document}